# Silicon-based Josephson junction field-effect transistors enabling cryogenic logic and quantum technologies

Yusheng Xiong, and Kaveh Delfanazari
*James Watt School of Engineering, University of Glasgow, Glasgow G12 8QQ, UK*
Dated: 26/10/2025, kaveh.delfanazari@glasgow.ac.uk

*Abstract*—The continuous miniaturisation of metal–oxide–semiconductor field-effect transistors (MOSFETs) from long- to short-channel architectures has advanced beyond the predictions of Moore's Law. Continued advances in semiconductor electronics, even near current scaling and performance boundaries under cryogenic conditions, are driving the development of innovative device paradigms that enable ultra–low-power and high-speed functionality. Among emerging candidates, the Josephson Junction Field-Effect Transistor (JJFET or JoFET) provides an alternative by integrating superconducting source and drain electrodes for efficient, phase-coherent operation at ultra-low temperatures. These hybrid devices have the potential to bridge conventional semiconductor electronics with cryogenic logic and quantum circuits, enabling energy-efficient and high-coherence signal processing across temperature domains. This review traces the evolution from Josephson junctions to field-effect transistors, emphasising the structural and functional innovations that underpin modern device scalability. The performance and material compatibility of JJFETs fabricated on Si, GaAs, and InGaAs substrates are analysed, alongside an assessment of their switching dynamics and material compatibility. Particular attention is given to superconductor–silicon–superconductor Josephson junctions as the active core of JJFET architectures. By unfolding more than four decades of experimental progress, this work highlights the promise of JJFETs as foundational building blocks for next-generation cryogenic logic and quantum electronic systems.

## I. Josephson Junction Early Development

The Josephson effect was first discovered by B. D. Josephson in 1962 [1], and further explanation was made on the phenomenon of tunnelling supercurrents [2], and phenomenological theory was developed [3]. In 1963, Anderson and Rowell experimentally observed an anomalous dc tunnelling current at or near zero voltage, which behaved similarly to the expectation of the Josephson current [4]. Also, they reported the effect of magnetic fields on the direct current in a number of tunnel junctions of different dimensions in the same year [5]. Josephson oscillations (alternative Josephson effects) were detected [6] [7], and an experiment aimed at observing photon emission upon the occurrence of alternating superconducting current was described in [8]. The above investigations mark the initial stage of the research into Josephson effects, where two superconducting electrodes are separated by a thin insulator layer [9]. The subsequent investigations showed that the Josephson effects can exist beyond the scope of superconductor-insulator-superconductor geometry [9]. Supercurrents can exist in a weak link of any physical nature between superconductors, including metals, semiconductors, and superconductors [10] [11].

Josephson dc current can be represented by the product of the critical current ($I_C$), which is the maximum supercurrent that can flow through the weak link, and the sine function of the phase difference ($\varphi = \varphi_1 - \varphi_2$) between a pair of superconductors separated by the weak link [2],

$$I_S = I_C \sin(\varphi).$$

The potential difference across the Josephson junction can be numerically calculated by taking the time-dependent derivation of the phase difference [2] [3],

$V = \frac{h}{2e}\frac{d\varphi}{dt}$, where $h$ and $e$ are Planck's constant and the electronic charge, respectively.

Cooper pair is an important term in understanding the dc Josephson effects. In a medium of metal lattice/ions carrying the positive charges, an electron carrying the negative charge can attract the ions in its vicinity, thereby potentially forcing the ions to move closer to the electron, leading to ion displacement and higher positive charge densities around this electron, overcoming repelling force existing between a pair of electrons and forcing them to pair up as a Cooper pair, known as electron-lattice vibration (phonon) interaction [12] [13].

Andreev reflection also explains the mechanism of the Cooper pair by demonstrating the process of charge transport at interfaces [14]. As demonstrated by Andreev, the range of the elementary excitations of the normal metal with both sides contacting the superconductors is quantised at energies not exceeding the energy gap of the superconductor [14] [15]. In stationary situations, the supercurrent across the Josephson junction is examined by microscopic theory [16] [17],

$$I_S(\varphi) \propto \int_{-\infty}^{\infty} dE [1 - 2f(E)] \, Im\{I_E(\varphi)\}.$$

In this expression, the supercurrent depends on the electron energy distribution function $f(E)$ and the spectral current $Im\{I_E(\varphi)\}$ characterised by the distance d between the superconductors and the transport parameters of the junction's material, including resistivities, Fermi velocities, and interface parameters [11].

## II. From CMOS to beyond-CMOS: Josephson junction FET

Field-Effect Transistors have been playing an indispensable role in many fields, including electronics, medical and telecommunication, for several decades since their invention. In FET, the current flowing through the channel between the drain and source terminals is controlled by the potential difference between the drain and source, and meanwhile is modulated by the gate voltage-induced electric field, as demonstrated in Figure 1. In today's modern world, silicon wafer-based CMOS technology has been dominating the chip fabrication for a broad range of electronic products. In the CMOS structure, as demonstrated in Figure 2, the voltage source is connected to the gate metal, inducing the field effect in the channel, which is beneath the gate oxide. Also, two voltage sources are applied to the separate doped regions located at the sides of the channel. By modulating the carrier concentration of either p- or n-type carriers in the channel, the current flowing through the channel is modulated for either analogue or digital circuitry applications, depending on how the CMOS transistors are combined in the circuits.

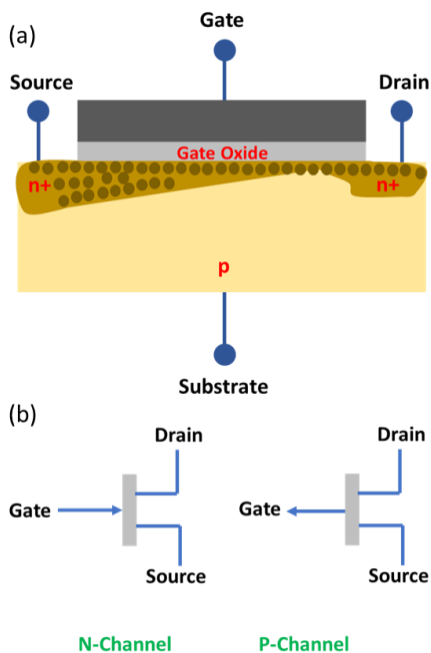

*Figure 1. Schematic of field-effect transistors (a) showing a p-type substrate with n-type doped source and drain areas. The carrier concentration in the channel beneath the gate oxide is modulated by the gate voltage, which induces the field effect through the gate oxide. Symbols of n-type channel field-effect transistors and p-type channel field-effect transistors (b) showing the three terminals: drain, source, and gate.*

The past sixty years have witnessed a transformative evolution in the development of FET technology. Gordon E. Moore made a statement in 1965 that the number of transistors that could be placed on a chip is doubling every year without increasing the cost of the chips [18]. The more devices on a smaller chip, the cost per device decreases and with reduced power consumption per component. Triggered by the COVID-19 pandemic, the rate of digitalisation of the world surged, enabled by the semiconductor industry and innovation procedure [19]. In 1975, the previous prediction was updated by Moore that the number of components on a single chip would double every two years, combining scaling component size and increasing chip area [20]. In 1965, chips of the minimum feature size of about 50 μm and of 50 components in total were produced in the industry, in contrast to the minimum feature size of 10nm on a single chip incorporating several billion transistors reported in 2017 [21].

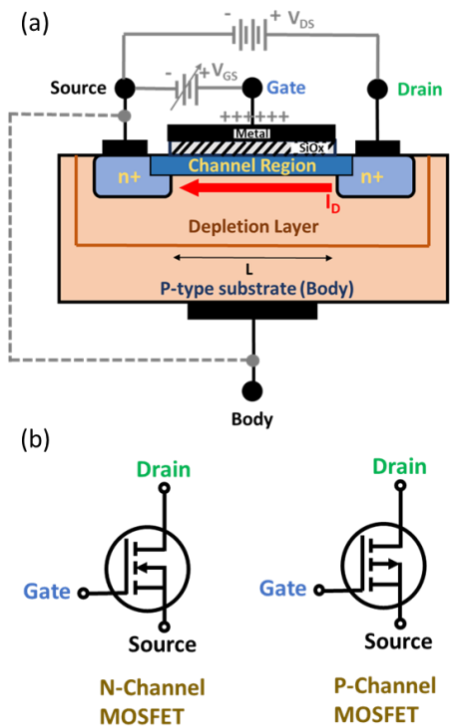

*Figure 2. Schematic of the fundamental metal-oxide-semiconductor field-effect transistors (a) showing the p-type substrate where the n-type doped drain and source regions are bridged by the channel beneath the silicon oxide. The carrier concentration in the channel is modulated by the voltage source applied to the metal gate. The source and drain electrodes are fabricated on the n-type doped region for electrical connection. Symbols of n-type and p-type MOSFET (b) showing the respective three device terminals: drain, source and gate.*

Scaling theory was proposed by Dennard et al. in 1974 [22] [23], marking the starting point of producing sophisticated technologies of electronic devices [24]. Scaling theory, referred to as 'classic' or 'traditional' scaling methodology [21], states that the transistor gate length, gate width, gate oxide thickness, and supply voltage are all scaled by the same scaling factor, while the channel doping is scaled by the inverse of the same scaling factor [23]. Such scaling methodology was achieving great success in the industry up until 130nm technology in the early 2000s [21]. In the past 20 years (until 2017), the doubling of transistor density every two years as predicted by the 1975 Moore's law was maintained by developing new generations of process technologies, while the general 14nm and 10nm technology had taken longer to develop due to increased process complexity and more photomasking steps [21].

The concept of 'More Moore' introduced in 2007 discusses the continued scaling of horizontal and vertical physical feature sizes of silicon-based complementary metal oxide semiconductor (CMOS) transistors [24] [25]. Moreover, numerous publications have analysed the scaling effects and predicted the future of CMOS development [24] [26] [27] [28] [29] [30] [31] [32] [33] [21] [34] [35] [36] [37]. In recent decades, beyond CMOS technologies, challenges and solutions have been discussed in [38] [39] [40] [41] [42] [43] [44] [45] [46]. In 2021, the 'Beyond CMOS' technology-related concepts corresponding to the new emerging materials and devices, device interactions and applications, device and material integrations, performance assessment, and difficult challenges were presented in the International Roadmap For Devices And Systems [47].

An outstanding document, 'Moore's Law – Now and in the Future', published in 2022 by the Intel Executive, releases the following information regarding the current and future situations of Moore's Law [19]:

- The predicted end of Moore's Law in the mid to late 2010s has been greatly exaggerated, and the pace of Moore's Law can be sustained through ongoing innovation. Intel aims to deliver around 1 trillion transistors in a single chip by 2030.
- Due to the scaling effects, the device sizes are decreasing, indicated by the transistor innovation over time, including the invention of Strained Silicon (Intel 90nm), Enhanced Strain (Intel 65nm), High-K Metal Gate (Intel 45nm), Enhanced HKMG (Intel 32nm), First FinFET (Intel 22nm), Enhanced FinFET (Intel 14nm), COAG (Intel 10nm), Enhanced FinFET with Super MIM Capacitor (Intel 10nm), and EUV Lithography (Intel 4nm).
- Intel's next architectural innovation is RibbonFET, coming with the implementation of the gate-all-around (GAA) transistor in the Angstrom era, with sizes down to 20A, and PowerVia with sizes down to 18A. RibbonFET represents the first new transistor architecture since FinFET. It is expected to achieve transistor performance per watt parity by leadership in 2025.
- The next generation of extreme ultraviolet (EUV) lithography, also referred to as High Numerical Aperture or 'High NA', can bring further improvements in the resolution as well as error reduction.
- Most importantly, in terms of the aims and objectives of this paper, in the meantime, the quantum realm is embraced, not only in the form of quantum computing, but also in investigating new concepts in physics and material science that are potentially in revolutionising the way computing is done. The pursuit of Moore's Law requires addressing the exponential growth in the power consumption of CMOS-based computation, and ultra-low-power solutions that exploit quantum effects at ambient room temperature are necessary.

This requirement for ultra-low power consumption aligns with the motivation for investigating high-Tc Josephson junction FETs. Furthermore, until today, a great amount of effort has been invested in the investigation of beyond-CMOS quantum computing and technologies, and cryo-CMOS incorporated quantum technologies [48] [49] [50] [51] [52] [53] [54] [55] [56] [57] [58] [59] [60] [61]. They demonstrate a potential in addressing the power consumption issues presented in the CMOS-based computation, providing a platform for the continuous pursuit of Moore's Law.

Promising candidates, superconducting three-terminal devices, operating based on the proximity effect induced by the weak link and similar in structure to MOSFETs, were proposed in [62] [63]. Other types of three-terminal superconducting devices have also been proposed in [64] [65] [66]. It was demonstrated in [67] that by the year 1979, it had yet to achieve a gate-tunable superconducting device for the space between the drain and source electrode larger than 1 μm [68]. The possibility of fabricating a voltage-controlled hybrid Josephson junction FET, as stated by the authors from [62], was first discussed in 1971 [69]. JJFET is supposed to consist of a superconductor-semiconductor-superconductor Josephson junction with the gate electrode deposited above [62]. The first type of JJFET is formed on an accumulation layer. The substrate with n-type doping provides the source of electrons. Under the positive gate voltage, more electrons are attracted to beneath the gate oxide, and thus the channel becomes an n+ type semiconductor, forming the accumulation layer. Therefore, the conductance of the accumulation layer as well as the critical current is controlled by the gate voltage. Another configuration is the inversion layer formed on a p-type substrate, where the drain and source are n+ doped, and the n-type channel beneath the gate oxide is formed by applying the positive gate voltage, converting the channel from p-type to n-type. Compared with the accumulation layer, the inversion layer can offer improved conductance controllability due to the minimised substrate leakage current. The junction current can be made dependent on the gate voltage at cryogenic temperatures, since a lightly-doped p-type substrate would be frozen out due to the finite ionisation energy of acceptors [70] and thus almost no hole mobility, increasing the dependency of the junction current on the gate voltage. MES JJFET utilises the depletion mode in the channel for operation. It comprises a substrate on the bottom, a buffer layer in the middle and an n-type epilayer on top serving as a channel which is in intimate contact with the source, drain and gate electrode. The n-type epilayer is formed by the ion-implantation technique. Different from the MOS JJFET mentioned above, the MES JJFET has no gate insulation; rather, it forms a depletion region beneath the gate electrode when the negative voltage is applied. This depletion region insulates the gate electrode from the epilayer. Besides, the thickness of the depletion region controlled by gate voltage modulates the channel resistance, therefore modulating the current flow in the channel. MES JJFET has higher transconductance than that of MOS JJFET [62]. Besides, MES JJFET has two characteristic features of depletion profiles. For a small drain-source voltage, the device can be made to pinch off; For a large drain-source voltage, the device can operate in the saturation region [62]. Among the superconducting three-terminal devices, Josephson junction field-effect transistors are one promising candidate among the competitors for the realisation of the cryogenic logic and analogue circuit applications with zero power consumption, taking advantage of their superconducting merits. The principles of Josephson junction field-effect transistor operation can be electrically equivalent to the Resistively-Capacitively-Shunted-Junction (RCSJ) model [71], where the resistance representing the channel resistance and the capacitance representing the gate oxide capacitance between the drain and source points are connected in parallel, as indicated in Figure 3 (a).

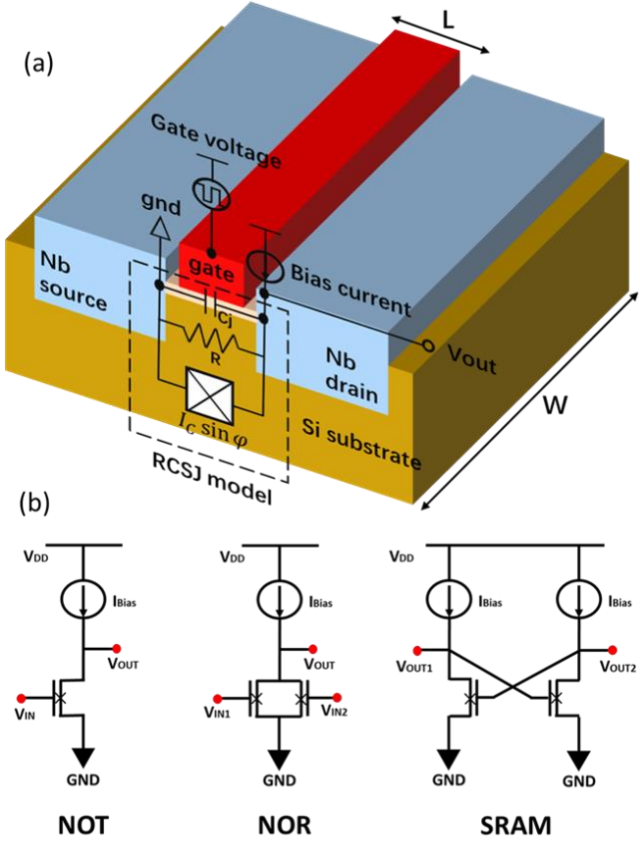

*Figure 3. (a) Schematic of Josephson junction field-effect transistor based on a silicon wafer showing the basic structures in terms of the substrate (dark yellow), superconductor niobium source and drain (blue), gate oxide (light pink), and the gate (red) on it. The electrical characteristics of the device are represented by the RCSJ model, where the channel resistance and gate oxide capacitance between the drain and source are connected in parallel. W and L stand for the device width and the gate length, respectively. (b) Schematic of Josephson junction field-effect transistor logic symbols for NOT gate, NOR gate, and SRAM (from left to right) [72].*

Meanwhile, another contributor to this electrical circuit is the phase-difference function between the drain and source electrodes. In the operation, one superconductor is grounded, while another is biased with the current source or voltage source to generate the current flowing through the channel, which is between the drain and source and underneath the gate oxide layer. Also, the voltage source applied to the gate can vary the current flowing in the channel by modulating the carrier concentration. The device can work in either the superconductive or non-superconductive (resistive) mode, depending on the comparison between the critical current and the drain current. The critical current value is the maximum allowed drain current for the superconductive state. When the drain current exceeds the critical current, the device works in resistive mode and generates a non-zero voltage output. Conversely, the device works in the superconductive mode and produces zero-voltage output with zero power consumption. The resistance of the 'normal' channel is independent of the voltage across the weak link. For a current-biased junction, the relationship between the voltage and current is given by [72]

$$V = R_N \sqrt{I^2 - I_0^2},$$

where $V$ is the voltage across the junction, $R_N$ is the normal resistance of the channel, $I$ is the supplied current, which has a Josephson current-phase relation with the critical current $I_0$,

$$I = I_0 \sin\theta,$$

as derived from the time-independent Ginzburg-Landau theory [73]. The critical current is related to the superconductor energy gap [72]:

$$I_0 = \frac{4\pi\Delta}{2eR_N}.$$

Furthermore, bringing the Josephson junction field-effect transistors into the realm of cryogenic digital logic and analogue circuits, which completely revolutionise the way computing is done in the conventional room temperature CMOS system, can potentially mean a transformative evolution towards nano-scale and even atomic-scale zero-power computation. As indicated in Figure 3 (b), Josephson field-effect transistors can compose elemental logic gates such as the NOT gate, NOR gate, and SRAM [72]. Different ways of combining them in the circuit can potentially achieve an increased level of flexibility, functionality, and complexity in the cryogenic conditions by exploiting their superconducting mechanisms. Figure 4 presents the development of experimentally realised JJFETs fabricated on silicon, GaAs, and more advanced InGaAs materials from the early 1980s until today.

III. SUPERCONDUCTOR-SILICON-SUPERCONDUCTOR JOSEPHSON JUNCTION

**Sandwich-type Josephson junction on a membrane etched from bulk silicon**
The first type of superconductor-silicon-superconductor is a sandwiched Josephson junction, which was realised by wet etching an opened area of a single-crystal silicon bulk from the top surface down to very close to its bottom surface, forming a locally thinned silicon junction barrier heavily doped by boron, followed by deposition of superconducting materials on both the top and bottom surfaces, forming the Josephson junction [74] [75] [76] [77] [78]. Several merits have been gained regarding this novel fabrication technique, including the controllable thickness of silicon barrier before superconductor deposition, a clean silicon crystal barrier, and the resulting low junction capacitance and high ratios of critical current to capacitance and high product of critical current and normal resistance, providing promising platforms for potential applications in a detector or mixer [74] [75]. They also suggest that the removal of silicon oxide can further improve the junction performance for high-frequency applications [75].

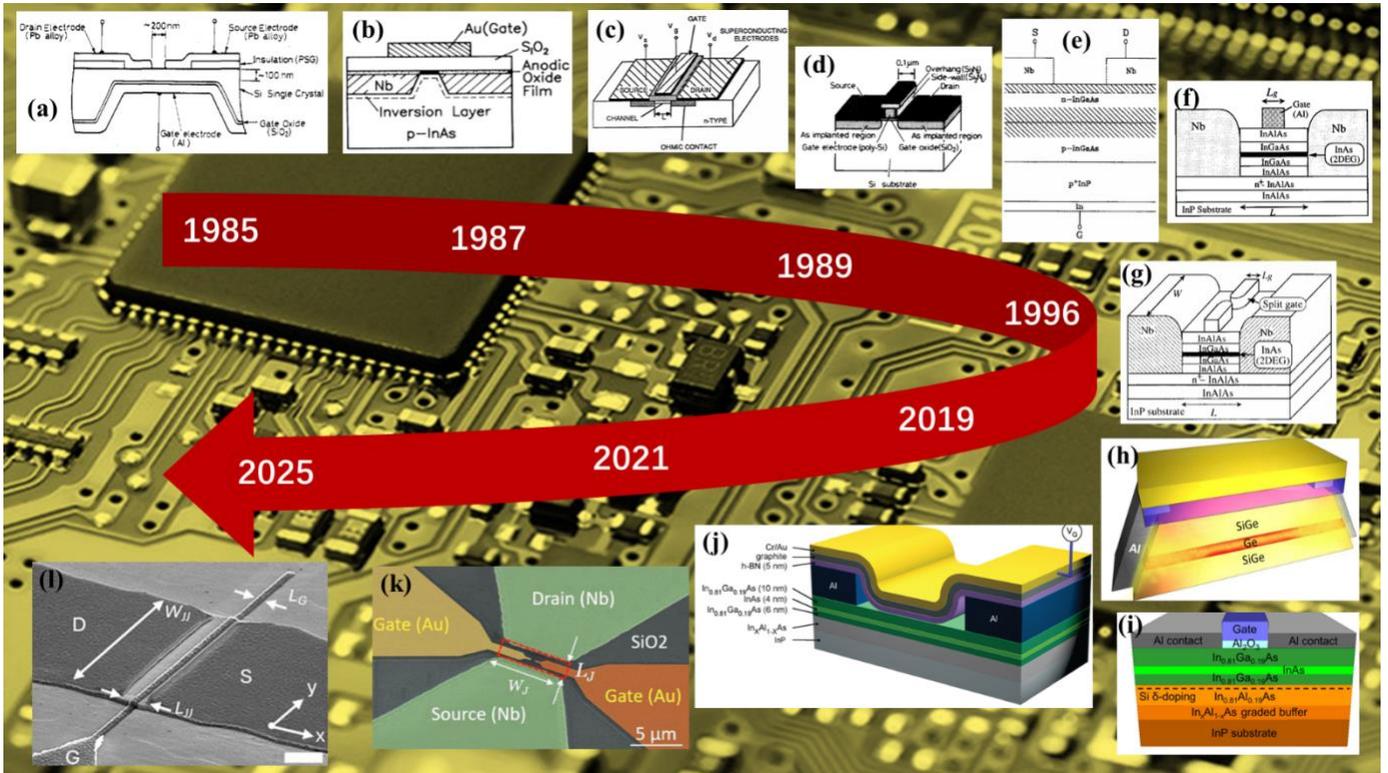

*Figure 4. Diagram showing the experimental development of gated Josephson junctions. (a) Cross-sectional view of a three-terminal superconducting device where the gate and drain/source electrode are fabricated on the opposite surface of the thinned silicon wafer [68]. (b) Schematic view of a gated InAs-substrate with a Nb-superconducting Josephson Junction and the gate electrode deposited on the junction [79]. (c) Schematic of a superconducting FET fabricated on a GaAs substrate [80]. (d) Schematic of Nb superconducting FET fabricated on a silicon wafer [81]. (e) Schematic of superconducting InGaAs junction field-effect transistors with gate, source, and drain terminals indicated [82]. (f) Schematic view of the structure of a Josephson junction FET with an InAs-inserted-channel inverted HEMT structure [83]. (g) Cross-sectional view of the structure of a split-gated Josephson junction on an InP substrate [84]. (h) Schematic of a germanium quantum-well Josephson field-effect transistor [85]. (i) Schematic of the InAs quantum well heterostructure JJFET[86]. (j) Schematic of epitaxial Al-InAs heterostructure JJFET showing the h-BN gate dielectric, aluminum superconducting leads, gate electrode, and InP substrate[87]. (k) False-colored SEM image of one gate-voltage-controlled hybrid symmetric and planar 2D junction [88]. (l) Tilted-view SEM micrograph of an InAsOI-based Josephson FET [89].*

As reported in [76], a sandwiched silicon-membrane device was fabricated using similar techniques from [74] [75], and low-shunt-capacitance electrode configurations were realised on the thin silicon membrane. As estimated from Figure 5 in [76], around 7mA critical current was observed for a 60nm chemically thinned silicon channel. An 80nm ultrathin Nb-Si-Nb silicon membrane was fabricated from bulk silicon [78]. Experimental results demonstrate the simultaneous occurrence of tunnel behaviour and Josephson coupling [78] [90], and possibly the first-time observation of multiple Andreev reflection in a silicon membrane [78].

It was observed that the superconductivity in silicon was induced by the proximity effect, which was dependent on carrier concentration [91]. The increase in carrier concentration in the silicon layer enhances the induced superconductivity in silicon due to the superconducting proximity effects [91]. A Nb-Si-Nb sandwich structure on single-crystal silicon was fabricated and measured, where the Kupriyanov-Lukichev theory for the proximity effect in SINIS structures was applied to describe the supercurrent [77]. Superconducting quantum wells were proposed to describe these double-barrier structures [77]. Similar aspects and discussions are also presented in [92].

**Coplanar Josephson junction on bulk silicon**

As described in [76], a coplanar superconductor-silicon-superconductor Josephson junction was fabricated by the following procedures. The surface and a depth of 0.5um of a polished bulk silicon was heavily doped by boron diffusion, followed by removal of boron glass layer by dipping in 10 percent HF solution, and by removal of silicon-boron phase layer by low-temperature oxidation and HF dip. It is suggested that to minimise the regrowing silicon oxide layer prior to superconductor lead deposition, the silicon wafer should be initially dipped in a 10 percent HF solution, followed by rinsing in $HNO_3$ solution at 100 ℃ for 5 minutes and another dip in HF solution for 2 minutes. After superconductor deposition, photolithography and electron beam lithography for the respective purposes of pad profile etching and junction gap etching completed the device fabrication. The coplanar lead-on-silicon junction, especially adaptable to array design, can perform both hysteretic and non-hysteretic V-I relationships. With slight changes in both the sandwich and coplanar structures, the behaviour of a super-Schottky diode performing high sensitivities suitable for low series resistance design can be obtained.

A coplanar Josephson junction device composed of two superconducting Pb-In alloy electrodes separated by a silicon bulk channel was fabricated [93] using a similar approach described in [76]. The silicon surface was heavily doped to p-type by boron diffusion, followed by the subsequent removal of the boron glass layer and the silicon-boron phase layer [93]. After cleaning of silicon surface, electron beam lithography and resist development were completed prior to sputtering of Pb-In alloy, and therefore, the junction gap was formed after removing the resist [93]. The same device was measured at 4.2K and 1.5K, respectively, to compare its V-I characteristics. The device became more hysterical as the temperature decreased. Besides, no devices fabricated on an undoped silicon surface showed evidence of supercurrent. In contrast, five out of 50 to 60 devices fabricated on doped silicon showed Josephson coupling, while all of them showed Josephson behaviour. It concludes a relatively high yield of super-Schottky devices, while a low yield of Josephson devices awaits further improvement in metal-semiconductor interfacial quality and gap definition [93].

It was reported in [90] [94] that differential conductance showed a peak at zero voltage at 4K for Nb-Si-Nb coplanar surface-contact geometry, although the zero-state resistance was not achieved at temperatures down to 1.2K. On the other hand, ridge contact geometry on bulk silicon demonstrates supercurrent at lower than 5K [90]. Furthermore, Nb-Si-Nb coplanar surface-contact Josephson junctions on n-type heavily doped silicon were also investigated [95]. Subgap structures were observed in V-I characteristics, which could be explained by multiple Andreev reflections, and the probability of occurrence of these reflections was determined by the barrier strength between the superconductor and semiconductor, which further determined the transmissivity of the barrier [95]. Larger barrier strength can lead to tunnelling behaviour with deficit current, while the smaller barrier strength can lead to metallic behaviour with excess current [95]. One of the S-Sm-S junctions behaving V-I characteristic, but without the observation of supercurrent, was explained by the quasi-particle theory of a leaky tunnel junction [95]. Attempts were made to achieve an irreversible shift from tunnelling behaviour to metallic behaviour by applying a large current or voltage [95]. The junction shifted to excess current from deficit current after the application of large current, and similar results were obtained by the application of large voltage or by thermal treatment of the interface at 600℃ for a few seconds [95]. No supercurrent was observed, however. Niobium silicide formation resulting from thermal treatment cannot effectively reduce the junction barrier height due to the very similar Schottky barrier height between the silicon and niobium, and the Schottky barrier height between the silicon and niobium silicide [95]. At high doping levels, the Schottky barrier height can be even higher due to the surface states, in contrast to the theory of WKB approximation, which becomes invalid in the high doping ranges [95].

Besides, silicene-based coplanar Josephson junctions were also investigated. It was observed that the 0-pi transition was induced by exchange splitting, and the critical supercurrent did not vanish at the transition point [96].

**Sandwich-type Josephson junction on amorphous silicon**

Another type of superconductor-silicon-superconductor Josephson junction is achieved by deposition of amorphous silicon (a-Si) on a layer of superconductor, the base electrode, followed by deposition of the second layer of superconductor, the counter electrode, on a-Si, forming a sandwiched vertical superconductor-a-Si-superconductor geometry. Nb-a-Si-Nb Josephson junction has been fabricated and investigated [97] [98] [99] [100] [101] [102] [103] [104].

Hydrogenated a-Si can be formed in a hydrogen ambient during the sputtering. Adding hydrogen was found to increase the Josephson current density and improve reproducibility. Furthermore, hydrogenated a-Si can be activated into n-type by the incorporation of $PH_3$ gas into the discharge or into p-type by the incorporation of $B_2H_4$ gas during the sputtering process. The Josephson current density of n-type a-Si was found to be considerably higher than that of p-type of the same thickness [97].

For the Nb-a-Si-Nb Josephson junction, which was sputter etch treated to remove the oxidised a-Si layer before Nb counter layer deposition, the Josephson current density varied exponentially over several orders of magnitude with barrier thickness [98]. A 2nm amorphous niobium phase sublayer between the a-Si barrier and Nb counter electrode layer, revealed by cross-sectional TEM, could considerably decrease the effective energy gap due to the proximity effect occurring in this sublayer [100]. The knee visible at around the energy gap voltage value in the V-I curve was another indication of the proximity effect occurring in the counter Nb electrode. The existence of the thin Nb sublayer was also confirmed by the voltage-conductance curve measured at 77K, where the curve minima deviated from the zero voltage, indicating non-symmetrical junction geometry [100]. It was demonstrated that Nb-a-Si-Nb junction properties did not degrade with annealing at temperatures up to 300 ℃ [101]. The experimental results indicate that such a device is a potential candidate for the application of higher speed and higher density superconducting circuits [101].

The effects of hydrogenation in a-Si barrier were investigated by comparing the characteristics of Nb-Nb Josephson junction sandwiching a sole un-hydrogenated a-Si barrier, a sole hydrogenated a-Si:H barrier, and a composite a-Si-a-Si:H-a-Si barrier layer [105]. The hydrogen was incorporated into a-Si layer by introducing $H_2$ plasma into Ar during the a-Si sputtering processes [105]. Experimental data showed that the $I_CR_N$ product of the composite a-Si-a-Si:H-a-Si barrier had a significantly higher value than that of the sole un-hydrogenated a-Si barrier [105]. It is concluded that the a-Si:H layer induced by a mix of Ar-$H_2$ plasma gases can lead to lower subgap current resulting from a lower density of localised states in a-Si:H layer [105]. Similarly, the effects of hydrogenation in a-Si barrier were investigated in Nb-a-Si-a-Si:H-a-Si-Nb Josephson junction where the hydrogenated a-Si:H layer was sandwiched by un-hydrogenated a-Si layers, and an abrupt improvement in measured characteristic was seen, indicating that the localized states near the geometric center of the barrier were most important in device characteristic degradation [106], similar to the conclusion from [105]. The effects of hydrogen concentration in a-Si on NbN-a-Si:H-Nb barrier height were investigated [107]. The average barrier height increased nearly linearly with hydrogen concentration in a-Si [107], indicating a reduction of surface states and dangling bonds within the barrier [107]. It is also demonstrated that the quality of the interface between Nb and barrier (Si cap-a-Si:H-Si cap) can be improved by replacing the barrier Si cap with a Ge cap [107]. Similar conclusions are presented in [108] regarding the relationships between the interface barrier height and hydrogen concentration in NbN-a-Si-a-Si:H-a-Si-Nb Josephson junction. Theory on resonant tunnelling related to the density of localised states was proposed and was in good agreement with experimental results [102] [109]. It is suggested that the experimental data can be demonstrated by the resonance mechanism of electrical charge transferring through the silicon interlayer related to the presence of localised states in the forbidden band of the a-Si [99].

Another similar edge-type geometry is a novel 'sandwiched' Nb-a-Si-Nb junction, differing from the previously mentioned types. In this configuration, a-Si interlayer and the counter layer of superconductor are sputtered on the edge of the base layer of superconductor [110]. It was discovered that the junction conductivity had an unusual dependence on the a-Si layer thickness, which was attributable to the resonant nature of the current passing through the interlayers having abrupt plane-parallel boundaries [110]. A similar step-edge type on the Nb-a-Si-Nb junction was also investigated [111]. Devices reported demonstrate promising application for the radiometer due to the properties of high sensitivity, good chemical and mechanical stability, and thermal recyclability [111].

Selective Niobium Anodization Process (SNAP), a novel approach to selectively anodise the upper counter niobium layer uncovered by a photoresist mask, was proposed to define the area of sandwich-type Nb-a-Si-Nb junction [112]. In this process, the anodised niobium layer can separate the upper niobium layer from the bottom niobium layer, forming a well-isolated area of junction with a diameter larger than 3um [112]. Another advantage is the inherent cleanliness of the process [112]. The fabrication of a sandwich junction employing such a method never requires removing the sample from the vacuum system until the junction sandwich layer deposition is completed [112]. Such a SNAP method was applied to NbN-a-Si-a-Si:H-a-Si-Nb sandwich junction fabrication [104] [107] [113] [108], Nb-a-Si-a-Si:H-a-Si-Nb sandwich junction fabrication [105] [106], Nb-a-Si-Nb sandwich junction fabrication [98] [99] [105] [114] [115], Nb-a-Si-Pb sandwich junction fabrication [99], Nb-a-Si:H-Nb sandwich junction fabrication [105], Nb-oxidised a-Si-Nb sandwich junction fabrication [116], NbN-a-Si-Nb sandwich junction fabrication [117] [114], and NbN-oxidised a-Si:H-NbN sandwich junction fabrication [118]. Furthermore, SNAP Nb-oxidised a-Si-Nb Josephson junction sandwich technology was brought to the field of logic gate application, benefiting from the advantages of SNAP processing, including inherent cleanliness, yielding superior junction quality, uniformity, and run-to-run reproducibility compared with conventional processing [116]. This technology shows promise for reliable, high-yield, large-scale integrated (LSI) Josephson circuits [116]. The suitability of SNAP-patterned Nb-a-Si-Nb tunnel junctions for LSI circuit applications was examined in [115], following the process for medium-scale integration described in [116]. The first aspect of the suitability examination was the improved precision of the junction area defined by the $SiO_x$ mask [115]. By replacing the photoresist mask used in the previous works with a $SiO_x$ mask, the undercut phenomena degrading the junction resolution could be effectively eliminated [115]. Besides, the examination of the uniformity of critical current density suggested that the improvement in photolithography and the use of RIE could provide increased control over the junction areas, resulting in smaller variations in critical current density [115]. Moreover, the junction excellence in stability manifested in their ability to withstand the lasting annealing temperatures higher than those processes in which they were fabricated [115]. Higher annealing temperatures could increase the critical current, while thermal cyclings between 4.2K and room temperature could have little effect on the critical current [115]. NbN-oxidised a-Si:H-NbN Josephson tunnel junctions were also applied to the chain circuits consisting of four-junction logic (4JL) gates with the minimum logic delay down to 9ps per gate, resulting from the reduction in junction capacitance [118].

A systematic investigation of the performance of oxidised a-Si layer as a tunnelling barrier was reported on various superconductors, including Nb, $Nb_3Sn$, and $V_3Si$ [119]. Junctions made with 2-3nm a-Si tunnelling barrier were expected to be the most reliable and of the highest quality due to a homogeneous oxide layer resulting from exposure to room air [119]. Amorphous Si oxidation time versus normal tunnelling resistance and critical current density was investigated in a high-stability NbN-a-Si-NbN sandwich Josephson junction prepared by vertical rf sputtering [120]. The normal tunnelling resistance increased exponentially with oxidation duration, while the critical current density showed a contrasting tendency [120]. The effects of sputter etch treatment on oxidised a-Si were investigated in [98]. Prior to deposition of the Nb counter electrode, rf-sputtered a-Si was exposed to air, growing an oxide layer [98]. One type of device was sputter etched to remove the oxide before Nb counter layer deposition, while another type was not [98]. By comparing the two types, the gap voltage fell from 2.8mV to 2.7mV, while the $I_CR_N$ product value fell from 1.4mV to 0.9mV, decreasing much more obviously than the gap voltage [98]. Also, the capacitance was reduced by a factor of five to six compared with the oxide barrier [98]. Additionally, the effects of oxidation on a-Si were investigated in a Nb-Al-a-Si-Pb Josephson junction, where the partially oxidised a-Si region functioned as a double-height barrier [121]. The a-Si layer was partially oxidised, forming an insulating layer as the tunnelling barrier, and it was found that the partially oxidised a-Si barrier could yield high-quality Josephson junctions and could guarantee fast operation of the device [121]. Besides, increasing the a-Si thickness could lead to degradation of tunnelling performances due to the intrinsic properties of the unoxidized a-Si fraction [121]. Double-layered sandwiched Nb-NbN-oxidised a-Si:H-NbN-Nb tunnelling junctions were investigated in [122]. It was reported that the value of $V_g$ and $V_m$ saturated at NbN thickness larger than 100nm [122]. By using the double-layered electrodes of 100nm NbN and 50nm Nb films, the effective penetration depth could be sufficiently decreased without losing the large gap voltage of 4.4mV and the small subgap parameter of 14mV [122]. The effects of a particular dopant, tungsten, were investigated in the Nb-a-Si-Nb junction [123]. Besides, different degrees of tungsten doping in the a-Si layer were also investigated. A fully degenerated a-Si layer indicated SNS-type conduction, which could be described by a classical resistive model, while a lower degree doped a-Si layer manifested a resonance mechanism of current transport. The properties of the Josephson junction could be modulated substantially by varying the impurity concentration [123].

A NbN-a-Si-NbN device was fabricated, and a great improvement was reported in the gap voltage, the product of the critical current and normal tunnelling resistance, the sub-gap quasi-particle current parameter, and the ratio of the thickness of a-Si to its relative dielectric constant [124]. By comparison with all-Nb junctions, the NbN/a-Si interface can have a larger barrier height than Nb/a-Si [117]. Similarly, the sandwiched type NbN-a-Si-Nb junction performance was investigated in [104]. It is reported that high-power sputter etch treatment on the NbN surface for contamination removal prior to a-Si sputtering can substantially reduce the gap voltage [104]. This demonstrates why the junction barrier prepared from a composite layer of hydrogenated a-Si sandwiched between a-Si layers, a-Si-a-Si:H-a-Si, can show even smaller critical current density compared with the standard barrier (a-Si only), if the base NbN electrode is treated by high-power sputter etch [104]. All-NbN Josephson tunnel junctions, where the a-Si:H barrier was surface oxidised, were fabricated and brought to the field of logic circuit applications [118]. High-speed operation was realised with the logic delay as low as 9ps per gate, demonstrated by the reduction in junction capacitance resulting from the thickness increase in a-Si barrier [118]. Since the high-quality NbN thin films were typically sputtered at an elevated temperature of 700℃, amorphous silicon barrier technology compatible with the high-quality NbN deposition at high temperatures was developed to deposit NbN, a-Si-a-Si:H-a-Si barrier, and Nb layer, composing the SNAP sandwich tunnelling junction [113]. Reduced NbN energy gap was observed for a-Si barrier deposited at high substrate temperatures, likely due to the interaction between NbN and Si [113]. Besides, the average barrier height as a function of barrier deposition temperature demonstrates a sharp increase in the average barrier height from 35mV to 95mV from below 400℃ to above 450℃ [113].

**Step-edge Josephson junction on bulk silicon**

A novel, but less commonly reported geometry of a superconductor-silicon-superconductor Josephson junction is the step-edge junction [125]. In this geometry, certain areas of the silicon bulk are ion-milling etched several hundred nanometers away, forming a sharp and vertical step, self-separating the superconductors, which are evaporated from an angle [125]. An obvious advantage is the ease in fabrication steps since the high-resolution lithography techniques are no longer required to separate the superconductors, as well as the junction gap [125]. On the other hand, the superconductor separation length is limited to higher than 100nm since the evaporated superconductors are likely to cover the entire step edge shorter than 100nm, leading to shorts of superconductors, therefore making such a technique challenging to fabricate junctions with lengths shorter than the coherence length [125]. The devices fabricated by step-edge techniques exhibited high resistance values, negligible junction capacitance due to non-overlapping superconductors, and ideal Josephson coupling [125]. Additionally, it is concluded that the coherence length in heavily doped silicon is approaching a factor of four higher than the value estimated using the relationship for the coherence length of normal metals [125].

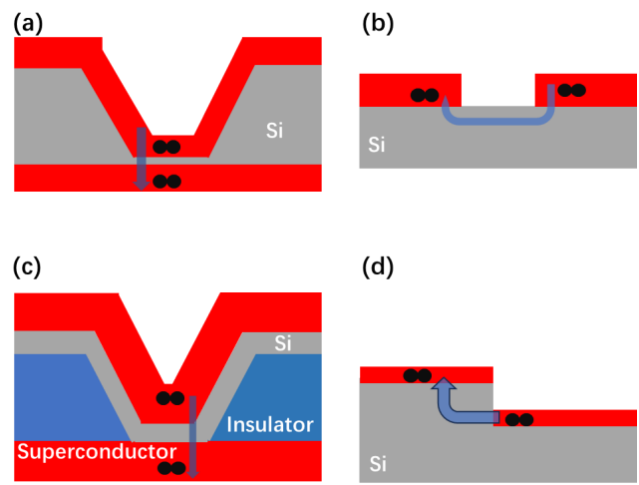

*Figure 5. Schematics showing four different configurations of superconductor-Si-superconductor Josephson junction, with (a) sandwiched superconductor-Si-superconductor Josephson junction, (b) surface contacted superconductor-Si-superconductor Josephson junction, (c) sandwiched superconductor-a-Si-superconductor Josephson junction, and (d) step-edge superconductor-Si-superconductor Josephson junction, with red, grey, and blue regions indicating the superconductor, Si and a-Si, and insulator, respectively. The black dots and blue arrow indicate the flow of Cooper pairs through the junction.*

Figure 5 visualises the four main configurations of superconductor-Si-superconductor Josephson junctions demonstrated in this section. Moreover, a review of the field of semiconductor-barrier Josephson junctions, earlier than the year 1980, is presented in [126].

IV. SUPERCONDUCTOR-SILICON-SUPERCONDUCTOR JOSEPHSON JUNCTION FIELD-EFFECT TRANSISTORS

Coplanar-type Josephson junction field-effect transistors based on single-crystal silicon as a semiconductor-coupled barrier were experimentally realised for the first time [68], with its device schematic indicated in Figure 4 (a). It was observed that the proximity-effect induced Josephson tunnelling current flowing between the source and drain superconducting electrodes could be controlled by gate bias voltage [68]. A threshold gate voltage of 50mV was required to trigger an observable critical current, as implied in Figure 4 in [68]. On the other hand, the device structure and gate voltage reported were yet to be optimised for higher compatibility with the requirements of digital circuits and systems applications [68]. Also, the gate electrode fabricated on the opposite side of the source and drain makes it challenging to apply in integrated circuits [81]. Planar devices composed of gate electrode, gate oxide, and superconductor electrodes were fabricated on the same side of the single-crystal silicon wafer, as reported in [81] shown in Figure 4 (d). However, challenges are raised in the fabrication since the gate electrode must sit between the source and drain [81], imposing higher requirements in the lithography alignment and junction dimension. Another significant difference reported in [81] is the gate-controllable normal-state resistance arising from the doping profile of $n^{++}$-p-$n^{++}$ of source-channel-drain, in contrast to $p^{++}$-$p^{+}$-$p^{++}$ doping profile reported in [68].

Besides the direct deposition of superconductor on silicon wafer, another approach is to form a silicon compound by evaporating non-superconducting metal on silicon wafer, followed by high-temperature treatment to form superconducting metal silicide. Compared with the direct deposition, metal silicide formed by high-temperature annealing can form a clean interface with the silicon channel. On the other hand, the transition temperature of metal silicide depends on its final thickness [127], annealing conditions [127], the type of metal [128], and the metal/silicon ratio [128] in the composition after formation. In the context of JJFET, platinum has been used to form platinum silicide on the Silicon-On-Insulator substrate by rapid thermal processing in the moderate temperature range (300 °C- 500 °C) [127]. However, the critical temperature of this superconductor has been recorded at 1.03K [127], far lower than that of the NbN superconductor of our interest in this paper discussion. Thin film A15 phase vanadium silicide has been synthesised by carefully regulated interdiffusion between the metallic vanadium and the underlying silicon layer, with a critical temperature higher than 13 K [129]. A bulk crystalline superconducting molybdenum silicide nanowire has been obtained by evaporation of molybdenum on the bulk silicon wafer, followed by rapid thermal annealing, with the superconducting critical temperature recorded ranging from 2.54K to 3.40K [130].

Besides the reported superconducting silicon JJFET, as shown in Figure 4 (a) and (d), progress has also been achieved in other materials. It has been approved in [79] that a superconducting proximity effect exists in the two-dimensional electron gas (2DEG) in the native inversion layer on p-type InAs substrate, as shown in Figure 4 (b). Superconducting source and drain can be weakly bridged through thinned ohmic contacts and accumulation layer, as shown in Figure 4 (c) [80]. Moreover, in [82] the results show that large voltages trigger normal FET operation while small voltages trigger Josephson junction or super-Schottky diode operation, with its schematic shown in Figure 4 (e). The merits of JJFET have been enhanced by using the high electron mobility transistor (HEMT) type gate, as shown in Figure 4 (f) [83]. Furthermore, a submicron split gate modulated 2DEG junction channel was investigated in terms of the superconducting and normal transport in the semiconductor-coupled Josephson junctions in the clean limit, with its vertical device layout shown in Figure 4 (g) [84]. SiGe/Ge/SiGe quantum well heterostructure was integrated with aluminum superconducting leads for JJFETs and superconducting quantum interference devices (SQUIDs) realisation, with its device schematic shown in Figure 4 (h) [85]. A systematic investigation including the gate voltage dependence of critical current, normal state conductance, and characteristic voltage has been reported for the InAs quantum well heterostructure JJFETs [86], as shown in Figure 4 (i). The effects of h-BN gate dielectric have been investigated in the epitaxial Al-InAs heterostructure JJFET, and the study shows that the h-BN dielectric modulates the channel density less compared with the aluminium oxide deposited by atomic layer deposition (ALD) [87], with its device schematic shown in Figure 4 (j). The first successful fabrication of a large array of chip-integrated hybrid JJFETs realised on Nb-InGaAs platform has been reported, and the systematic investigation of conductance switching performance under the gate voltage has been demonstrated in [88], with the false-colored SEM device image shown in Figure 4 (k). InAs on Insulator (InAsOI) has also been demonstrated as a promising platform for developing hybrid JJs and JJFETs with the combined advantages of utilising high-k gate dielectric, including aluminium oxide and hafnium oxide [89], with the device image shown in Figure 4 (l). These experimental demonstrations on Si, GaAs, InGaAs, InP, and Ge-based JJFETs imply promising prospects in the development of energy-efficient quantum systems, and new emerging technologies in terms of the cryogenic Boolean logic and analogue circuitry and even artificial-intelligence (AI) potentially integrable systems.

Finally, Table 1 is included in the appendix to summarise over 40 papers on the experimental demonstration of the superconductor-Si-superconductor Josephson junction and the JJFET based on this junction, with the figure-of-merit parameters recorded and compared. Some figure-of-merit parameters are mentioned across these publications in the table. Gap voltage, $V_g$, indicates the sum of the energy gaps; the subgap leakage parameter, $V_m$, is a general indicator of the junction's suitability for digital applications, since the higher the $V_m$, the more current to be switched out of the junction into the load; The product of critical current, $I_C$, and channel normal resistance, $R_N$, is another important measure of junction quality; the capacitance per unit area influences the response time of the junction for both the digital and mm-wave applications [117].

V. CONCLUSION

This review has outlined the evolution from Josephson junctions to modern field-effect architectures, highlighting the emergence of the JJFET as a key enabler of cryogenic and quantum-compatible logic. As semiconductor scaling approaches its physical and power dissipation limits, JJFETs provide a promising route to ultra–low-power, high-speed, and phase-coherent operation. Comparative analyses of superconductor–semiconductor junctions across Si, GaAs, InGaAs, and Ge platforms reveal consistent progress in junction quality, switching speed, and material integration. Key performance metrics, including the supercurrent magnitude and tunability, gap voltage, subgap leakage, and $I_cR_n$ product, serve as critical benchmarks for evaluating device scalability, junction quality, and the feasibility of implementing gate-controlled logic and quantum switching operations. Continued advances in interface engineering, epitaxial growth, and nanoscale fabrication will be essential to realise JJFETs as foundational elements for hybrid quantum–classical systems, bridging superconducting logic, cryogenic electronics, and next-generation quantum technologies.

**Appendix: Table 1**

Table summarising the literature in experimentally reported superconductor-Si-superconductor Josephson junction and the JJFETs based on superconductor-Si-superconductor Josephson junction. Information recorded in this table includes the silicon wafer specification, base superconductor, counter superconductor, junction length, junction width, critical current, junction normal resistance, barrier dopant type and concentration, the product of critical current and normal resistance, gap voltage, subgap leakage parameter, junction capacitance, cryogenic measurement temperature, the year of publication, reference, gate type, gate insulator and thickness, and gate material and thickness.

| Specs | Base superconductor | Counter superconductor | $L_{JJ}$ [nm] | $W_{JJ}$ [um] or junction area [um²] | $I_C$ [uA] | $R_{NN}$ [ohm] | Barrier dopant type and concentration | $I_C R_{NN}$ product | Gap voltage ($V_g$) | Subgap leakage parameter ($V_m = I_c R_{sg}$) | Junction capacitance | T [K] | Year | Reference | Gate type | Gate insulator & thickness | Gate material & thickness |
|---|---|---|---|---|---|---|---|---|---|---|---|---|---|---|---|---|---|
| P-type single-crystal <100> silicon wafer | Pb/In/Au | Pb/In/Au | 200 | 20um | 140uA at 180mV gate voltage (estimated from fig. 4) | NA | Boron, 5E18cm⁻³ | NA | NA | NA | NA | 4.2K | 1985 | [68] | Back | 40nm, SiOx | 700nm, Al |
| Single-crystal silicon wafer | Nb | Nb | 150 | 60um | 32uA at 3.2V gate voltage (estimated from fig. 4) | NA | Boron, 5E22m⁻³ | 0.08mV-0.3mV | NA | NA | NA | 4.2K | 1989 | [81] | Top | 10nm, SiOx | Poly-Si |
| P-type SOI | PtSi | PtSi | 30-70 | 2.5um | 3.2uA | 1E4-1E6ohm | Boron, 1E19cm⁻³ | NA | NA | NA | NA | 150mK | 2021 | [127] | Top | 3.5nm, SiOx | NA |
| Single-crystal <111> p-type silicon wafer | Pb In | Pb In | NA | NA |  | NA | Boron, 6.5E18cm⁻³ – 2.3E20cm⁻³ | NA | NA | NA | NA | 1.5K | 1970 | [131] |  |  |  |
| Sandwich type on single-crystal <100> p-type silicon wafer | Indium | Indium | 40 | 87.5 | 33.6mA (theoretical maximum) | 0.025 | Boron, 7E19cm⁻³ | NA | NA | NA | NA | 3.3K | 1974 | [74] |  |  |  |
| Sandwich type on <100> single-crystal silicon wafer | NA | NA | 125 | Junction area (140um by 140um) | 31.5mA at 1.9K | 0.026ohm at 4.2K | Boron, 7E19cm⁻³ | NA | NA | NA | 15.6pF | 1.9K | 1975 | [75] |  |  |  |
| Coplanar type on <100> single-crystal silicon wafer | Lead | Lead | 100 and 300 | 30 | 1mA and 10mA | No more than 0.1 | Boron, 2E20cm⁻³ | NA | NA | NA | NA | 1.9K and 4.2K | 1977 | [76] |  |  |  |
| Sandwich type on <100> single-crystal silicon wafer |  |  | 60 | Junction area (50um²) | 7mA (estimated from figure 5) | 0.55ohm | Boron, 1.2E20cm⁻³ | 2.15mV | NA | NA | NA | 4.2K |  |  |  |  |  |
| Sandwich type on phosphorus doped a-Si:H | Nb | Nb | 6.5 | Junction area (25um by 50um) | Exceed 1000A/cm² | NA | Phosphorus, doped by PH₃ discharge | 1.1mV highest, 0.8mV for most | NA | NA | NA | 4.2K | 1979 | [97] |  |  |  |
| Sandwich type on oxidized a-Si | Nb, Nb₃Sn, V₃Si | Pb | 0-20 | Junction area (2×10⁻³cm²) | 1-10A/cm² | 10³ohm/cm² for 2nm a-Si (predicted from figure 2) | Undoped | NA | NA | NA | 2.8±0.3uF/cm² | 4.2K | 1980 | [119] |  |  |  |
| Sandwich type on a-Si:H | NbN | NbN | 3-21 | Junction area (10um by 10um) | 1000A/cm² | NA | Undoped | 1.6mV-2.2mV | 4.4mV | 3mV-13mV | Dielectric constant of 9.3 | 4.2K | 1981 | [124] |  |  |  |
| Sandwich type on a-Si:H | Nb | Nb | 7 | Larger than 3um in diameter | 25A/cm² | NA | Undoped | NA | NA | 12mV | NA | 4.2K | 1981 | [112] |  |  |  |
| Coplanar type on degenerately doped <100> single-crystal silicon wafer | Pb-In alloy | Pb-In alloy | 100-300 | 20 | 3mA (estimated from figure 2(b)) | 3.3ohm | Boron, E20cm⁻³ | Up to 800uV at 4.2K | NA | NA | NA | 4.2K and 1.6K | 1981 | [93] |  |  |  |
| Sandwich type on oxidized a-Si | NbN | NbN | 5 | Junction area (40um by 50um) | 1000A/cm² (estimated from figure 3) | 10⁻⁶ohm.cm² (estimated from figure 3) | Undoped | 2.2mV | 4.4mV | NA | NA | 4.2K | 1981 | [120] |  |  |  |
| Sandwich type on a-Si | Nb | Nb | 4-6 | Junction area (4um by 4um) | 900A/cm² | NA | Undoped | 1.4mV | 2.8mV | 1.35mV and 1.45mV | 0.024pF/um² | 4.2K | 1982 | [98] |  |  |  |
| a-Si | NbN | Nb | 4-5 | NA | 500A/cm² | NA | Undoped | 1-1.3mV | 3.5mV | 15-22mV | 0.04pF/um² | 4.2K | 1982 | [117] |  |  |  |
| Oxidized a-Si:H | Nb/NbN | NbN/Nb | 3 | Junction area (10um by 10um) | 1000A/cm² | NA | Undoped | NA | 4.4mV | 14mV | NA | 4.2K | 1982 | [122] |  |  |  |
| a-Si | Ni Au | Al Al | 10 | 1.7×10⁻²cm | NA | NA | Undoped | NA | NA | NA | NA | 1.2K | 1982 | [132] |  |  |  |
| a-Si | Nb | Nb | 5-7 | Junction area (12um by 12um) | 110A/cm², 1400A/cm² | NA | Undoped | NA | NA | 11mV | 2.5uF/cm² (20% uncertainty) | NA | 1982 | [114] |  |  |  |
|  | NbN | Nb | 4-5 | NA |  |  |  |  | 3.85mV | Above 20mV | NA |  |  |  |  |  |  |
| Oxidized a-Si | Nb | Nb | NA | Junction area (12um by 12um) | 110A/cm², 1400A/cm² | NA | Undoped | NA | NA | 9mV-11mV | 2.5uF/cm² (20% uncertainty) | 4.2K | 1983 | [116] |  |  |  |
| a-Si-a-Si:H-a-Si | Nb | Nb | 5.6 | NA | NA (not extractable from figure 2) | NA | Undoped | 1.6mV | NA | 28mV | NA | 4.2K | 1983 | [105] |  |  |  |
| Oxidized a-Si-a-Si:H-a-Si | NbN | Nb | 5-7 | 3E-7cm² | 0.3 – 10A/cm² | Several hundred ohms | Undoped | 1.2mV | 3.2mV | 35mV | NA | 4.2K | 1983 | [104] |  |  |  |
| a-Si | Nb | Nb | NA | Junction area (12um by 12um, 7um by 7um, 4um by 4um) | NA | NA | Undoped | NA | NA | 6mV-12mV (extracted from figure 5) | NA | 4.2K | 1983 | [115] |  |  |  |
| Oxidized a-Si:H | Nb-NbN | NbN | 1-2 | Junction area (2.5um by 2.5um, 10um by 10um) | 1.75kA/cm² | NA | Undoped | 2.3mV | 4.3mV | 16mV | 4uF/cm² | 4.2K | 1984 | [118] |  |  |  |
| a-Si-a-Si:H-a-Si | Nb | Nb | 5.6 | NA | 1 – 100A/cm² | NA | Undoped | NA | NA | 10mV-28mV (estimated from figure 2) | NA | 4K | 1985 | [106] |  |  |  |
| (Oxidized a-Si)-a-Si-(Oxidized a-Si) a-Si-(Oxidized a-Si) | Nb | Pb | 6.5 3.9 | 5E-4cm² upper trace, 5E-10cm² lower trace | NA | NA | Undoped | NA | NA | NA | NA | 1.5K | 1985 | [102] |  |  |  |
| Al-oxidized a-Si | Nb | Pb | 3.5 | Junction area (0.4E-3cm²-1.5E- | 0.2mA | 0.004ohm.cm² | Undoped | NA | NA | 20mV-40mV at 4.2K | 2uF/cm² | 4.2K | 1986 | [121] |  |  |  |

| Structure | Electrode 1 | Electrode 2 | Barrier thickness (nm) | Junction size | Current density | Normal resistance | Doping | $V_g$ | $I_cR_n$ | Sub-gap | Capacitance | Temperature | Year | Ref |
|---|---|---|---|---|---|---|---|---|---|---|---|---|---|---|
| Step edge on n-type <111> single-crystal silicon wafer | Pb | Pb | 100-500 | 3-7 | 0.94mA | 1.25 | Phosphorus, 3E20cm$^{-3}$ | 1.18mV | NA | NA | negligible | 4.02K | 1986 | [125] |
| a-Si | Nb | Pb | 2.5-8.5 | Junction area (10um$^2$-100um$^2$) | 10A/cm$^2$-10$^5$A/cm$^2$ (estimated from figure 3) | (5-5000) ohm.um$^2$ | Undoped | 0.3mV-1mV | NA | NA | 0.013pF/um$^2$ | 2.2K | 1987 | [99] |
| Sandwich type on <100> single-crystal silicon wafer | Pb alloy | Al | 50 | NA | NA | NA | Boron, 1E25m$^{-3}$ and 8E25m$^{-3}$ | NA | NA | NA | NA | 4.2K | 1987 | [91] |
| Sandwich type on a-Si-a-Si:H-a-Si | NbN | Nb | 5 | Junction area (10um$^2$-500um$^2$) | 1600A/cm$^2$ for 20% Si:H hydrogen concentration | NA | Undoped | NA | 4.2mV | 51mV at 4.2K | NA | 2.1K | 1987 | [107] |
| Sandwich type on a-Si-a-Si:H-a-Si | NbN | Nb | 5 | Junction area (9um by 9um) | Exceeds 1000A/cm$^2$ | 1.8ohm | Undoped | NA | 4mV | 32mV | NA | 4.2K | 1987 | [108] |
| Edge type on a-Si | Nb | Nb | 3-15 | 1-4 | 0.1mA (estimated from figure 2) | 0-70ohm (estimated from figure 4a) | Undoped | 0.02mV-2mV (estimated from figure 4b) | 1.8mV | NA | NA | 4.2K | 1988 | [110] |
| Sandwich type on a-Si-a-Si:H-a-Si | NbN | Nb | 16 | NA | NA | NA | Undoped | NA | 3.8mV (150 °C) and 3.2mV (700 °C) | 29.3mV and 16.8mV for 150°C and 700°C deposition temperature respectively | NA | 4.2K | 1989 | [113] |
| Sandwich type on oxidized a-Si-a-Si | Nb | Nb | 6.3 | Junction area (5um by 5um) | 1000A/cm$^2$ | NA | Undoped | 0.9mV-1.0mV | NA | NA | NA | 1989 | [109] |
| Sandwich type on 2-inch double-side polished single-crystal <100> silicon wafer | Nb | Nb | 80 | Membrane area (100um by 100um) | 0.01mA (estimated from figure 2. Inset) | NA | Boron, 2E20cm$^{-3}$-3E20cm$^{-3}$ | NA | NA | NA | NA | 1.2K | 1991 | [78] |
| Coplanar surface-contact type on single-crystal silicon | Nb | Nb | 100 | NA | NA | NA | N-type doped, 1E20cm$^{-3}$ | NA | NA | NA | NA | 1.2K | 1991 | [90] [94] |
| Ridge contact on single-crystal silicon | Nb | Nb | 100 | NA | 350uA | NA | N-type doped, 1E20cm$^{-3}$ | NA | NA | NA | NA | 1.75K | | |
| Sandwich type on single-crystal silicon membrane | Nb | Nb | 80 | 500um$^2$ | 23uA | NA | P-type doped, 1E20cm$^{-3}$ | NA | 2.6mV | NA | NA | 1.2K | | |
| Edge-type on Nb-doped Si interlayer | Nb | Nb | 3-4 | Junction area (0.2um$^2$-1um$^2$) | NA | NA | Nb doped | NA | NA | NA | NA | 4.2K | 1991 | [103] |
| Step-edge type on a-Si | Nb | Nb | 3-15 | Junction area (0.2um$^2$-0.4um$^2$) | 1uA-200uA | From unit to hundred ohms | Undoped | Up to 1mV | 2.0mV | NA | 10$^{-14}$ F | 4.2K | 1991 | [111] |
| Sandwich type on single-crystal silicon wafer | Nb | Nb | 40-60 | Junction area (2um$^2$-600um$^2$) | 150uA-200uA (estimated from figure 3) | NA | Boron, 7E19cm$^{-3}$ | NA | 2.9mV | NA | NA | 1.2K | 1993 | [77] [92] |
| Coplanar surface-contact type on single-crystal silicon | Nb | Nb | Less than 80nm | 12 | 0.3uA-50uA | 33 | Phosphorus, 5E26m$^{-3}$ | Most 50uV | Around 3mV | NA | NA | 4.2K | 1995 | [95] |
| Sandwich type on a-Si | Nb | Nb | 10 | Junction area (20um by 20um) | 20uA at 4.2K | 15 | Undoped | NA | 1mV-2mV | NA | NA | 4.2K-10K | 1997 | [100] |
| Sandwich type on doped a-Si | Nb | Nb | 7-9 | Junction area (36um^2 or 81um^2) | 1.6mA highest (estimated from figure 8a) | NA | Tungsten, concentration around 11% | 0.3mV, 0.12mV | 1mV | NA | NA | 4.2K | 2012 | [123] |
| Sandwich type on doped a-Si | Nb | Nb | 100 | NA | 3mA/um$^2$ | NA | Nb doped | 1.39mV | NA | NA | 332fF/um$^2$ | 4K | 2023 | [101] |